\documentclass[aps,prd,floats,preprintnumbers,nofootinbib]{revtex4}
\usepackage{epsfig,graphicx,color,amsmath}

\begin{document}

\preprint{NUHEP-PH/10-25}

\title{Low-scale Leptogenesis and Dark Matter}

\author{\vspace{0.5cm} Wei-Chih Huang$\,^{a,b}$}
\affiliation{\vspace{0.5cm}
\mbox{$^{a}$Department of Physics and Astronomy, Northwestern University,
Evanston, IL 60208}\\
 \mbox{$^{b}$High Energy Physics Division, Argonne National Laboratory,
Argonne, IL 60439}
\vspace{0.8cm}
}

\pacs{14.60.Pq, 14.60.St}

\begin{abstract}
The addition of gauge singlet fermions to the Standard Model Lagrangian renders the neutrinos massive and allows one to explain all that is experimentally known about neutrino masses and lepton mixing. At the same time, the gauge singlet fermion decays in the early universe produce a lepton asymmetry, which is converted to a baryon asymmetry via Spharelon processes (leptogenesis). On the other hand, the addition of a gauge singlet scalar to the Standard Model yields a thermal dark matter candidate through interactions between the Higgs boson and the gauge singlet scalar. By imposing a $Z_2$ symmetry on the gauge singlet scalar and one of the gauge singlet fermions, we can have viable dark matter candidates and new interactions coupling the $Z_2$-odd scalar to the $Z_2$-odd fermion, which can lower the leptogenesis scale (and the reheating temperature) to $\mathcal{O}$(TeV).

\end{abstract}

\maketitle

\section{Introduction}
Gauge singlet fields are a simple but very interesting form of physics beyond the Standard Model (SM). A gauge singlet scalar($S$) can be a thermal dark matter candidate by having $S$ couple to the Higgs boson (for example, see \cite{McDonald:1993ex}). In order to have it be stable or long-lived, a symmetry might be imposed or any decay channel has to be suppressed by a high energy scale, usually comparable to the Planck scale.

On the other hand, gauge-singlet fermions, known as right-handed (RH) neutrinos, can explain the observed tiny neutrino masses \cite{neutrino masses} via the type-I seesaw mechanism \cite{type-I seesaw}. RH neutrinos can also accommodate the observed baryon asymmetry \cite{baryon assymetry} via thermal leptogenesis \cite{first leptogenesis}. The mechanism of leptogenesis satisfies the three Sakharov's conditions \cite{sakharov} (i) baryon number violation, (ii) $C$ and $CP$ violation, (iii) deviation from thermal equilibrium. A lepton number asymmetry is generated via the decay of heavy RH Majorana neutrinos, which is converted into the baryon number asymmetry through Spharelon processes \cite{sphaleron}. In order to be the main source of the baryon asymmetry, the mass scale of the heavy RH neutrinos must typically be larger than $10^9$ GeV \cite{Davidson:2002qv}, which requires a high reheating temperature. Such a high reheating temperature leads to gravitino overproduction \cite{Bolz:2000fu} in the context of supersymmetry. There are many ways to avoid gravitino overproduction. Resonant leptogenesis \cite{resonant leptogenesis}, for instance, assumes the limit $m_{N_2}-m_{N_1}\ll m_{N_2}$ so that $m_{N_1}$ and $m_{N_2}$  can be as low as of order TeV. There is enhancement to the lepton asymmetry by taking into account flavor effects \cite{flavor effect}.
In \cite{Buchmuller:2005eh}, \cite{Davidson:2008bu}, and references therein, there are more detailed discussions on the solutions to the gravitino overproduction problem.

It is intriguing to combine these different ideas together, i.e., to have the gauge singlet scalar and fermions at once in the theory. This is done, e.g., in \cite{Petraki:2007gq}, which aims at lifting the tension between X-ray bounds and the Lyman-$\alpha$ bounds in the Dodelson-Widrow (DW) model \cite{DW model}, where the dark matter is the RH neutrino that is generated from neutrino oscillations. The constraint from X-rays puts an upper bound on the mass of the RH neutrino since it can decay into an active neutrino and a photon via loop diagrams and the decay rate is proportional to the mass of the RH neutrino. At the same time, Lyman-$\alpha$ puts a lower bound on the thermal-averaged momentum of the RH neutrino which equivalently is a lower bound on its mass. For more details, see \cite{Petraki:2007gq}. In the context of the DW model, the two constraints can not be satisfied at once. To reduce the tension between the two bounds, \cite{Petraki:2007gq} has both gauge-singlet scalar $S$ and gauge-singlet fermions $N$, where $S$ couples to $N$. $S$ decaying into $N$ provides an extra production mechanism of $N$ in addition to neutrino oscillations. $N$'s from the decay of $S$ get red-shifted when the universe cools down; therefore, they have a lower thermal average momentum than those generated by oscillations. Hence the Lyman-$\alpha$ bounds become weaker.
In \cite{Law:2007jk}, a similar setup is employed. With the help of an unbroken Abelian family symmetry, $G_{\textrm{family}}$, under which the full Lagrangian is invariant, all particles carry the charge of $G_{\rm{family}}$. The gauge singlet scalar, $S$, couples to gauge singlet fermions, $N$'s, with a nontrivial structure and, at the same time, the vacuum expectation value of $S$ provides masses to $N$'s,
in addition to the Majorana mass terms, in such a way that the Maki-Nakagawa-Sakata (MNS) matrix (or the Pontecorvo- Maki-Nakagawa-Sakata (PMNS, or MNSP) matrix) yields $\theta_{13}=0$ and some regions of the parameter space with the weak washout effect, which can yield thermal leptogenesis.

In this paper, we manage to add more structure into theory, i.e., an extra discrete symmetry, so that the framework can provide dark matter candidates, alleviate the gravitino problem of leptogenesis and make a connection between the baryon asymmetry and dark matter.
To be more specific, we introduce a gauge-singlet scalar $S$ and several gauge-singlet fermions $N$ and impose a $Z_2$ symmetry on $S$ and one of $N's$.
 In this situation, the lighter particle of $S$ and $N$ charged under the $Z_2$ symmetry can be the dark matter and the $Z_2$ symmetry guarantees the stability of the dark matter candidate. For leptogenesis, loop-diagrams with gauge singlets running inside give rise to the required strong and weak $CP$ phases for the generation of the lepton asymmetry. We found that the contributions to the lepton asymmetry from the new interactions can be a dominant source of matter-antimatter asymmetry without the problem of gravitino overproduction. \cite{McDonald:2007ka} uses a similar concept with the type-II seesaw mechanism.

This paper is organized as follows. In Sec.\ref{sec:formalism}, we describe the formalism and the particle content. In Sec.\ref{sec:leptogenesis}, a detailed analysis of leptogenesis from new interactions has been displayed. The discussion on dark matter is in Sec.\ref{sec:dark matter candidate} and we conclude in Sec. \ref{sec:conclusions}.

\section{the formalism}\label{sec:formalism}

The full Lagrangian can be written in the following way,

\begin{equation}\label{lagrangian}
\mathcal{L}=\mathcal{L}_{SM}+\mathcal{L}_1+\mathcal{L}_2+\mathcal{L}_3+\mathcal{L}_4,
\end{equation}
where
\begin{eqnarray}
  \mathcal{L}_1 &=& i\bar{N}_i \gamma^{\mu}\partial_{\mu}N_i-\frac{M_i}{2}\bar{N}_i^c N_i
  -y_{i\alpha}H \bar{N}_i L_j+ h.c.,\\
   \mathcal{L}_2 &=&i\bar{N} \gamma^{\mu}\partial_{\mu}N-\frac{M}{2}\bar{N}^c N,\\
  \mathcal{L}_3 &=&-V(H,S)=m^2_H|H|^2-\lambda_H|H|^4-\frac{1}{2}m_s^2S^2-\frac{1}{4}\lambda_SS^4
  -\frac{\lambda_{HS}}{2}|H|^2S^2,\\
  \mathcal{L}_4&=&-\lambda_i S \bar{N}_i^c\frac{1-\gamma_5}{2}N+h.c..
\end{eqnarray}
 $\mathcal{L}_{SM}$ refers to the SM Lagrangian and $S$ is a real singlet, whose vacuum expectation value (VEV) is zero, and its mass $m^2_S=m^2_s+\lambda_{HS}\langle H\rangle^2$. We introduce RH Majorana neutrinos, $N_i$, which generate tiny masses for active neutrinos, and $N$, which provides a new mechanism for leptogenesis. Here, we are interested in the situation with $3$ active, $3$ RH neutrinos, $N_i$, and one extra gauge-singlet Majorana fermion, $N$. The Lagrangian preserves a discrete symmetry under which $S\rightarrow -S$ and $N\rightarrow -N$ and the other particles remain unchanged. Therefore, there are no cubic terms for $S$ and no $yH \bar{L}_{\alpha}N$, i.e., $N$ is not responsible for the see-saw mechanism.

\section{Leptogenesis}\label{sec:leptogenesis}
\subsection{Baryon and lepton number}
We start by discussing how large the lepton asymmetry, $\epsilon$, should be in order to explain the observed baryon number asymmetry.
In \cite{Buchmuller:2005eh} \cite{Harvey:1990qw}, the relation between baryon ($B$) and lepton number ($L$) before and after sphaleron conversion is
\begin{equation}\label{B-L conversion via sphaleron}
    B(t_f)=c_s(B-L)=-c_sL(t_i),
\end{equation}
where subscripts, $i$ and $f$, refer to before and after the sphaleron process, respectively, $c_s=(8N_f + 4N_H)/(22N_f + 13N_H),$\footnote{In the SM, $c_s=28/79$ and $c_s=8/23$ for the MSSM.} in which $N_f$ is the number of generations, $N_H$ is the the number of Higgs doublets and $B-L$ is anomaly-free\footnote{In the context of the SM with additional RH neutrinos, the number of RH neutrinos for anomaly cancelation has to be three if $B-L$ is gauged. In this paper, $B-L$ is a global symmetry and there is no constraint on the number of RH neutrinos.} so that we do not need to assign a time index to it.
From the Wilkinson Microwave Anisotropy Probe (WMAP) data \cite{Hinshaw:2008kr},
\begin{equation}\label{wmap baryon photon ration}
    \eta_B=\frac{n_b}{n_{\gamma}}=\eta_{10}10^{-10}=273.9\, \Omega_b h^2\,10^{-10}=(6.23\pm 0.17)\times10^{-10}.
\end{equation}
 A similar quantity to $\eta$ is $Y$, which is the number density in a comoving volume, defined as \begin{equation}\label{definition of Y }
    Y=\frac{n}{s},
\end{equation}
where $n$ is the number density and $s$ is the entropy density. Obviously, the relation between $\eta$ and $Y$ is,
\begin{equation}\label{eta Y relation}
    Y_B=\eta_B\frac{n_{\gamma}}{s}.
\end{equation}

Assuming the entropy and baryon number per comoving volume remain constant from $t_f$ to the time of BBN, $t_{\rm{BBN}}$ and to that of matter-photon decoupling, $t_{\rm{WMAP}}$, then $Y_B(t_f)=Y_B(t_{\rm{BBN}})=Y_B(t_{\rm{WMAP}})$. Given that at $t_{\rm{WMAP}}$, $s=7.04\,n_{\gamma}$, we have
 \begin{equation}\label{Y B at tf}
   Y_B(t_f)=\eta_B\frac{n_{\gamma}}{s}=8.8\times10^{-11},
\end{equation}

and
\begin{equation}\label{eta L Y L at ti}
  Y_L(t_i)=-\frac{1}{c_s}Y_B(t_f)=-2.53\times10^{-10},
\end{equation} where we assume $N_H=2$ and $N_f=3$ for the MSSM.

\subsection{leptogenesis with new interactions}

\begin{figure}
\includegraphics[width=0.6\textwidth]{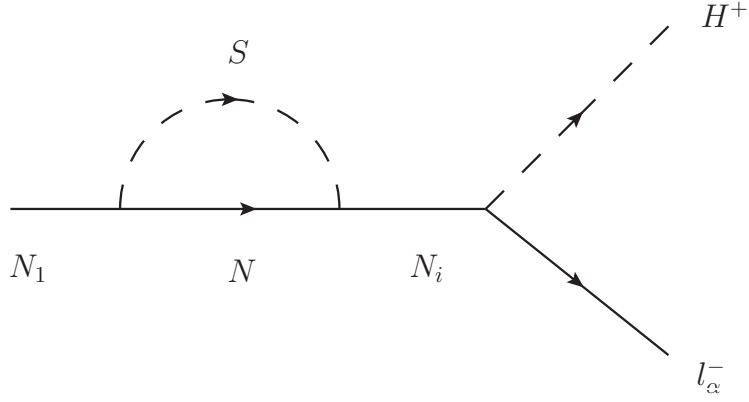}
\caption{Contribution to leptogenesis from new interactions.}
\label{fig:leptogenesis sunset diagram}
\end{figure}

\begin{figure}
\includegraphics[width=0.6\textwidth]{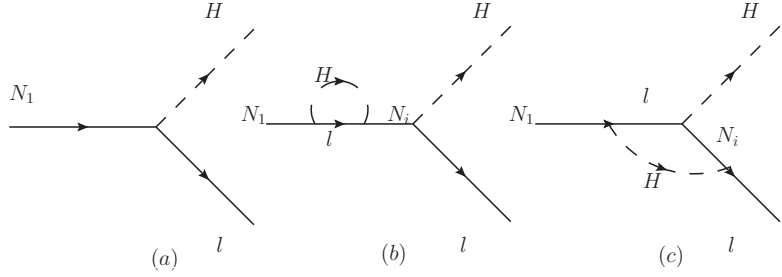}
\caption{Original leptogenesis.}
\label{fig:original leptogenesis diagram}
\end{figure}

 From the new interactions shown in Fig. \ref{fig:leptogenesis sunset diagram}, and the original seesaw ones in Fig. \ref{fig:original leptogenesis diagram}, we have

\begin{equation}\label{gamma1-gamma2}
    \Gamma(N_1\rightarrow H^+l^-)-\Gamma(N_1\rightarrow H^-l^+)=\left(-\sum_{i\neq1}\sum_{\alpha}
    \frac{\mathrm{Im}(y^*_{1\alpha}y_{i\alpha}\lambda_i\lambda^*_1)}{128\pi^2}
    \frac{m^3_{N_1}}{m^2_{N_i}-m^2_{N_1}}\right)+\left( \sum_{i\neq1} \frac{m_{N_1}\mathrm{Im}(yy^{\dagger})^2_{1i}}{64 \pi^2}f\left(\frac{m^2_{N_i}}{m^2_{N_1}}\right) \right),
\end{equation}where $m_{N_i}$ is the mass of $N_i$ ($i=1,2,3$) and, from \cite{f function definition},
\begin{equation}\label{f in leptogenesis}
    f(x)=\sqrt{x}\left(\frac{2-x}{1-x}-(1+x)\mathrm{ln}(\frac{1+x}{x})\right)\rightarrow -\frac{3}{2\sqrt{x}} \mbox{(when ${x\gg 1}$)}
\end{equation}
and we do not consider the degenerate situation, in which $m_{N_j}-m_{N_i}\ll m_{N_j}$.\footnote{Even without mass degeneracy, the self-energy diagram, (b) in Fig. \ref{fig:original leptogenesis diagram}, contributes to the lepton asymmetry. See \cite{f function definition} for more detail.}

 To simplify the expressions, we use the Casas-Ibarra parametrization \cite{Casas:2001sr}, which separates the high energy physics from the low energy one, and follow the procedure from \cite{Davidson:2002qv}, where they have obtained the bound of $m_{N_1}>10^9$ GeV for the standard type-I seesaw.
First, some quantities are defined as follows,
\begin{eqnarray}\label{definition of C-I parametrization}
  D_{\sqrt{m_N}} &=& \left(
                       \begin{array}{ccc}
                         \sqrt{m_{N_1}} & 0 & 0 \\
                         0 & \sqrt{m_{N_2}} & 0 \\
                         0 & 0 & \sqrt{m_{N_3}} \\
                       \end{array}
                     \right),
   \notag\\
    D_{\sqrt{m_{\nu}}} &=& \left(
                       \begin{array}{ccc}
                         \sqrt{m_{\nu_1}} & 0 & 0 \\
                         0 & \sqrt{m_{\nu_2}} & 0 \\
                         0 & 0 & \sqrt{m_{\nu_3}} \\
                       \end{array}
                     \right),\notag \\
  RR^t &=& R^tR=1_{3\times3},\notag\\
  y_{i\alpha}&=&(\frac{1}{\langle H\rangle}D_{\sqrt{m_{M}}}RD_{\sqrt{m_{\nu}}}U^{\dagger})_{i\alpha},
\end{eqnarray}
where $m_{\nu_i}$ are the active neutrino masses with $m_{\nu_3}\geq m_{\nu_2}\geq m_{\nu_1}$ and $U$ is the MNS matrix.

For pedagogical purposes, we repeat the derivation of the Davidson-Ibarra bound on $\epsilon_1$ with corrected coefficients for the original leptogenesis \cite{Davidson:2002qv}. From the second term in Eq. (\ref{gamma1-gamma2}), we obtain
\begin{eqnarray}\label{exercise of lower bound M1}
   \Gamma(N_1\rightarrow H^+l^-)-\Gamma(N_1\rightarrow H^-l^+)
   &=&
   \left( \sum_{i\neq 1} \frac{m_{N_1}}{64 \pi^2}\mathrm{Im}(yy^{\dagger})^2_{1i}f\left(\frac{m^2_{N_i}}{m^2_{N_1}}\right) \right), \notag\\
   &\simeq& \left( -\frac{3 m_{N_1}}{128 \pi^2}\sum_{i\neq 1} \mathrm{Im}(yy^{\dagger})^2_{1i}\frac{m_{N_1}}{m_{N_i}} \right), \notag\\
   &=&  \left( -\frac{3 m^2_{N_1}}{128 \pi^2}\sum_i \mathrm{Im}(yy^{\dagger}\frac{1}{D_{m_N}}y^*y^T)_{11} \right), \notag\\
   &=& -\frac{3}{128\pi^2\langle H \rangle^4}\sum_{\alpha} m^3_{N_1}m^2_{\nu_{\alpha}}\mathrm{Im}(R^2_{1\alpha}),
\end{eqnarray}
where we have used Eq. (\ref{definition of C-I parametrization}) and the fact that $(yy^{\dagger})_{11}$ is real. We have,

\begin{equation}\label{exercise of upper bound on epsilon}
    \epsilon_1=\frac{\Gamma(N_1\rightarrow H^+l^-)-\Gamma(N_1\rightarrow H^-l^+)}
    {\Gamma(N_1\rightarrow H^+l^-)+\Gamma(N_1\rightarrow H^-l^+)}=-\frac{3}{16\pi}\frac{m_{N_1}}{\langle H \rangle^2}
    \frac{\sum_{\alpha} m^2_{\nu_{\alpha}}\mathrm{Im}(R^2_{1\alpha})}{\sum_{\alpha} m_{\nu_\alpha} | R_{1\alpha}|^2}.
\end{equation}
For simplicity, we assume that light neutrino masses are hierarchical, i.e., $m_{\nu_3}\gg  m_{\nu_2}$, $m_{\nu_1}$,

\begin{eqnarray}\label{approximation on exercise of upper bound on epsilon}
  |\epsilon_1| &\leq&
  \frac{3}{16\pi}\frac{m_{N_1}}{\langle H \rangle^2}
    \frac{ m^2_{\nu_3}|\mathrm{Im}(R^2_{13})|}{ m_{\nu_3} | R_{13}|^2},  \\
   & \leq & \frac{3}{16\pi}\frac{m_{N_1}m_{\nu_3}}{\langle H \rangle^2},
\end{eqnarray} where equality holds when $|\mathrm{Im}(R_{13})|\gg|\mathrm{Re}(R_{13})|$. Thus, we obtain

\begin{equation}
    m_{N_1}\geq \frac{16\pi \langle H \rangle^2}{3 m_{\nu_3}} |\epsilon_1| \simeq 10^9 \mbox{GeV},
\end{equation} where we use the fact that the observed baryon asymmetry is due to leptogenesis to infer
 $|\epsilon_1| $. We will discuss this in more detail later.

Now, we turn to the new contribution to $\epsilon_1$ from new interactions, the first term in Eq. (\ref{gamma1-gamma2}). $\lambda_i$ are assumed real for simplicity, and
\begin{equation}\label{y yi}
   \sum_{\alpha}\mathrm{Im}(y^*_{1\alpha}y_{i\alpha})=\sum_{\alpha}\frac{\sqrt{m_{N_i}m_{N_1}}m_{\nu_{\alpha}}}{\langle H\rangle^2}
   \mathrm{Im}(R^*_{1\alpha}R_{i\alpha}).
\end{equation}

 Therefore, the total lepton number asymmetry from the decay of $N_1$ is,
\begin{eqnarray}\label{lepton asymmtry definition and result}
  \epsilon &=& \frac{\Gamma(N_1\rightarrow H^+l^-)-\Gamma(N_1\rightarrow H^-l^+)}
  {\Gamma(N_1\rightarrow H^+l^-)+\Gamma(N_1\rightarrow H^-l^+)
  +\Gamma(N_{1}\rightarrow N S)}, \\
   &=& -\frac{1}{\lambda_1^2+(yy^{\dagger})_{11}}
   \left(\sum_i\sum_{\alpha}\frac{\lambda_1\lambda_i}{16\pi}\frac{m^2_{N_1}}{m^2_{N_i}-m^2_{N_1}}
    \frac{\sqrt{m_{N_i}m_{N_1}}m_{\nu_{\alpha}}}{\langle H\rangle^2}\mathrm{Im}(R_{i\alpha} R^*_{1\alpha})
    +\frac{3}{16\pi\langle H \rangle^4}\sum_{\alpha} m^2_{N_1}m^2_{\nu_{\alpha}}\mathrm{Im}(R^2_{1\alpha})
    \right), \notag
\end{eqnarray}
where we use Eq. (\ref{gamma1-gamma2}) , (\ref{exercise of lower bound M1}), and (\ref{y yi}).

Note that $Y_L=\epsilon\eta' Y^{eq}_{N_1}$ ($T\gg m_{N_1}$) \cite{Giudice:2003jh}, where $\eta'$ is an efficiency factor which measures the wash-out effect and $Y^{eq}_{N_1}(T\gg m_{N_1})=135\zeta(3)/(4 \pi^4 g_*)$, where $g_*$ is the number of relativistic degrees of freedom in thermal equilibrium, $g_*\simeq 230$ for the MSSM. From \cite{Davidson:2008bu}, $\eta'\sim m_*/\tilde{m}_{\alpha\alpha}$ in the strong wash out scenario when $\tilde{m}>m_*$ and $\tilde{m}_{\alpha\alpha}>m_*$, where $m_* \sim 10^{-3}$ eV and
$\tilde{m}\equiv \sum_{\alpha} \tilde{m}_{\alpha\alpha}\equiv
\sum_{\alpha} 8\pi \frac{\langle H \rangle^2}{m^2_{N_1}}\Gamma(N_1\rightarrow H^+l^-, H^-l^+)=
\frac{\langle H \rangle^2}{m_{N_1}}(\lambda_1^2+(yy^{\dagger})_{11})$.
To sum, we have
\begin{equation}\label{master formula}
   Y_L \simeq \frac{135\zeta(3)}{4 \pi^4 g_*}\frac{m_*m_{N_1}}{\langle H\rangle^2}\frac{\left(
    \sum_i\sum_{\alpha}\frac{\lambda_1\lambda_i}{16\pi}\frac{m^2_{N_1}}{m^2_{N_i}-m^2_{N_1}}
    \frac{\sqrt{m_{N_i}m_{N_1}}m_{\nu_{\alpha}}}{\langle H\rangle^2}\mathrm{Im}(R_{i\alpha} R^*_{1\alpha})
    +\frac{3}{16\pi\langle H \rangle^4}\sum_{\alpha} m^2_{N_1}m^2_{\nu_{\alpha}}\mathrm{Im}(R^2_{1\alpha})
    \right)}
    {(\lambda_1^2+(yy^{\dagger})_{11})^2},
\end{equation}where we approximate $\tilde{m}_{\alpha\alpha}$ by $\tilde{m}$.

If $\lambda_i=0$, $\eta'=1$, $g_*=230$, $c_s=8/23$, and $m_{\nu_3}=\sqrt{\Delta m^2_{23}}$ in the MSSM and requiring the observed baryon asymmetry coming from leptogenesis, we have

\begin{equation}\label{reduce to davidson's result}
     Y_L(t_i)=\frac{8.8\times 10^{-11}}{c_s}
     \leq \frac{135\zeta(3)}{4 \pi^4 g_*}\frac{3}{16\pi}\frac{m_{N_1}m_{\nu_3}}{\langle H\rangle^2},
\end{equation}
which infers $m_{N_1}\geq 10^9$ GeV, that is consistent with the result in \cite{Davidson:2002qv}.

From Eq. (\ref{master formula}), it is obvious that $m_{N_1}$ could be smaller than $10^9$ GeV if one were to increase $\lambda_i$ or $R_{i\alpha}$.\footnote{Increasing $\lambda_1$ or $R_{1\alpha}$ would not allow one to reduce $m_{N_1}$, since they will be canceled by the denominator in Eq. (\ref{master formula}).}
From now on, we focus on the case with only two RH neutrinos, $N_1$ and $N_2$, for simplicity. The generalization to more RH neutrinos is straightforward.

Increasing $\lambda_2(\gg (yy^{\dagger})_{22})$ will cause $N_2$ to depart from thermal equilibrium at later times and the relic density of $N_2$ is roughly $\exp(-m_{N_2}/T_{D})*n_{\gamma}$ at decoupling, where $T_D$ is the decoupling temperature of $N_2$ and is fully determined by $\lambda_2$. However, as long as $m_{N_2}\gg m_{N_1}$ and $\lambda_2$ is large enough, the relic density of $N_2$ is too low to have any impact on the lepton asymmetry created by $m_{N_1}$ at $T\simeq m_{m_1}$.
 For instance, we choose $m_{N_2}=10 m_{N_1}$, $\lambda^2_1 \simeq (yy^{\dagger})_{11} \gtrsim \frac{m_* m_{N_1}}{\langle H\rangle^2}$,\footnote{This choice implies the strong wash-out scenario.} and $g_*=230$,
     \begin{eqnarray*}
       \frac{8.8\times 10^{-11}}{c_s} & \leq &  \frac{135\zeta(3)}{4 \pi^4 g_*} \frac{m_*m_{N_1}}{\langle H\rangle^2}
       \frac{1}{64 \pi \lambda_1^2} \frac{\lambda_2}{\lambda_1} \frac{m^2_{N_1}}{m^2_{N_2}-m^2_{N_1}}
       \frac{\sqrt{m_{N_2}m_{N_1}}m_{\nu_{\alpha}}}{\langle H\rangle^2}\mathrm{Im}(R_{i\alpha} R^*_{1\alpha}),\\
        &\leq & \frac{135\zeta(3)}{4 \pi^4 g_*} \frac{m_*m_{N_1}}{\langle H\rangle^2}
       \frac{\langle H\rangle^2}{64 \pi m_*m_{N_1}} \frac{\lambda_2}{\lambda_1} \frac{m^2_{N_1}}{m^2_{N_2}-m^2_{N_1}}
       \frac{\sqrt{m_{N_2}m_{N_1}}m_{\nu_{\alpha}}}{\langle H\rangle^2}\mathrm{Im}(R_{i\alpha} R^*_{1\alpha}).
     \end{eqnarray*}
Then, it is easy to show,
      \begin{equation}\label{increaing lambda2}
        5.4\times 10^{11}\leq \left(\frac{\lambda_2}{\lambda_1}\right) \left(\frac{m_{N_1}}{\mbox{GeV}}\right).
      \end{equation}
In a dramatic situation, we can have $m_{N_1}$ around $O(10\mbox{TeV})$ by having $\lambda_1\sim 10^{-7}$ and $\lambda_2\sim 1$.

 On the other hand, increasing $R_{2\alpha}$ yields larger $\epsilon$ without changing $\eta'$, which is independent of $R_{2\alpha}$.
  However, the estimate $\eta'\sim m_*/m_{\alpha\alpha}$, takes into account the effect of inverse decay($l+H\rightarrow N_1$) only. By increasing $R_{2\alpha}$, $\Delta L=2$ ($l^-H^+\leftrightarrow l^+H^-$ and $l^+l^-\leftrightarrow H^- H^+$) interactions will become important as well. To be more precise, from Eq. (\ref{master formula}), for MSSM in the limit of $m_{N_2}\gg m_{N_1}$ and strong wash out region, $\lambda^2_2\sim\lambda^2_1\geq(yy^{\dagger})_{11}\geq \frac{m_* m_{N_1}}{\langle H\rangle^2}$, the constraint on $m_{N_1}$ is
  \begin{equation}\label{r2 alpha case}
    10^{9} \leq \left( \frac{ m_{N_1}}{\mbox{GeV}} \right)\frac{m^2_{N_1}}{m^2_{N_2}}|\mathrm{Im}(R_{2\alpha}R^*_{1\alpha})|.
  \end{equation}
  For example, if we would like to make $m_{N_1}=10^8$ GeV, $\mathrm{Im}(R_{2\alpha})$ has to be $10\frac{m^2_{N_2}}{m^2_{N_1}}\sim 10^3$ for $m_{N_2}=10m_{N_1}$ and $|(R_{1\alpha})|\sim 1$. The ratio of $\Delta L=2$ interactions mediated by $N_2$ to those mediated by $N_1$ will be roughly $|R^2_{2\alpha}/R^2_{1\alpha}|^2\sim 10^{12}$, which is extremely large and has to be carefully considered in the calculation of $\eta'$. In other words, increasing $R_{2\alpha}$ may not be an efficient and applicable way to lower $m_{N_1}$.

\section{the dark matter}\label{sec:dark matter candidate}

As mentioned before, due to the discrete $Z_2$ symmetry, the lightest of $S$ and $N$ can be a thermal relic.
We first calculate the relic abundance for each of them, respectively, and then discuss if any of them can be the dark matter and at the same time the low-scale leptogenesis is viable.

\subsection{$S$ as dark matter($m_N > m_S$)}
From \cite{Kolb:1988aj}, we know the relic density of the dark matter is determined by the annihilation rate into SM particles in thermal equilibrium at the time of decoupling. A rule of thumb is that it decouples when the interaction rates with SM particles are smaller than the expansion rate of the universe. To be more quantitatively precise, the Boltzmann equation should be used (see chapter 5 in \cite{Kolb:1988aj}). With the help of \cite{McDonald:1993ex}, where a complex scalar is assumed, we can calculate $\langle \sigma v\rangle$ for $SS\rightarrow HH$
, $SS\rightarrow W^+W^-$, $SS\rightarrow ZZ$ and $SS\rightarrow f^+ f^-$, respectively.

$\bullet$$SS\rightarrow HH$
\begin{equation}\label{SS to HH}
\langle \sigma v\rangle=\frac{\lambda^2_{HS}}{128\pi m^2_S}\left(1-\frac{m^2_h}{m^2_S}\right)^{1/2},
\end{equation}

$\bullet$$SS\rightarrow W^+W^-$
\begin{equation}\label{SS to WW}
\langle \sigma v\rangle=\left(1+\frac{1}{2}\left(1-\frac{2m^2_S}{m^2_W}\right)^2\right)
\frac{\lambda^2_{HS}m^4_W}{8 \pi m^2_S((4m^2_S-m^2_h)^2+m^2_h\Gamma^2_h)}
\left(1-\frac{m^2_W}{m^2_S}\right)^{1/2},
\end{equation}

$\bullet$$SS\rightarrow ZZ$
\begin{equation}\label{SS to ZZ}
\langle \sigma v\rangle=\left(1+\frac{1}{2}\left(1-\frac{2m^2_S}{m^2_Z}\right)^2\right)
\frac{\lambda^2_{HS}m^4_Z}{16 \pi m^2_S((4m^2_S-m^2_h)^2+m^2_h\Gamma^2_h)}
\left(1-\frac{m^2_Z}{m^2_S}\right)^{1/2},
\end{equation}

$\bullet$$SS\rightarrow f^+f^-$
\begin{equation}\label{SS to ff}
\langle \sigma v\rangle=\frac{m^2_W}{\pi g^2}
\frac{\lambda^2_{HS}\lambda^2_f}{((4m^2_S-m^2_h)^2+m^2_h\Gamma^2_h)}
\left(1-\frac{m^2_f}{m^2_S}\right)^{3/2},
\end{equation}
where we have an extra factor $1/2$ compared to \cite{McDonald:1993ex} due to the fact that S is a real scalar field in our case.
The relic density of $S$ is given by (see Eq. ($2.7$) in \cite{McDonald:1993ex}),

\begin{equation}\label{relic density S}
    \Omega_S=\frac{\rho_S}{\rho_c}\frac{g(T_{\gamma})}{g(T_{fs})}
    \frac{K}{T_{\gamma}x_{fs}\langle \sigma v\rangle}
    \frac{T^4_{\gamma}}{\rho_c}
    \frac{1-\frac{3 x_{fs}}{2}}{1-\frac{x_{fs}}{2}},
\end{equation}
 where $\rho_S$ is the energy density of $S$, $\rho_c$ is the critical energy density of the universe, $T_{fs}$ is the freeze-out temperature for $S$, $x_{fs}=T_{fs}/m_S$, $T_{\gamma}$ is the present photon temperature, $g(T)$ is the the number of relativistic degrees of freedom around temperature $T$, and $K=
 (4\pi^3 g(T)/45M^2_{pl})^{1/2}$.
 An extra factor $1/2$ is again because of $S$ being a real scalar.

\subsection{$N$ as dark matter($m_S > m_N$)}
\begin{figure}
\includegraphics[width=0.6\textwidth]{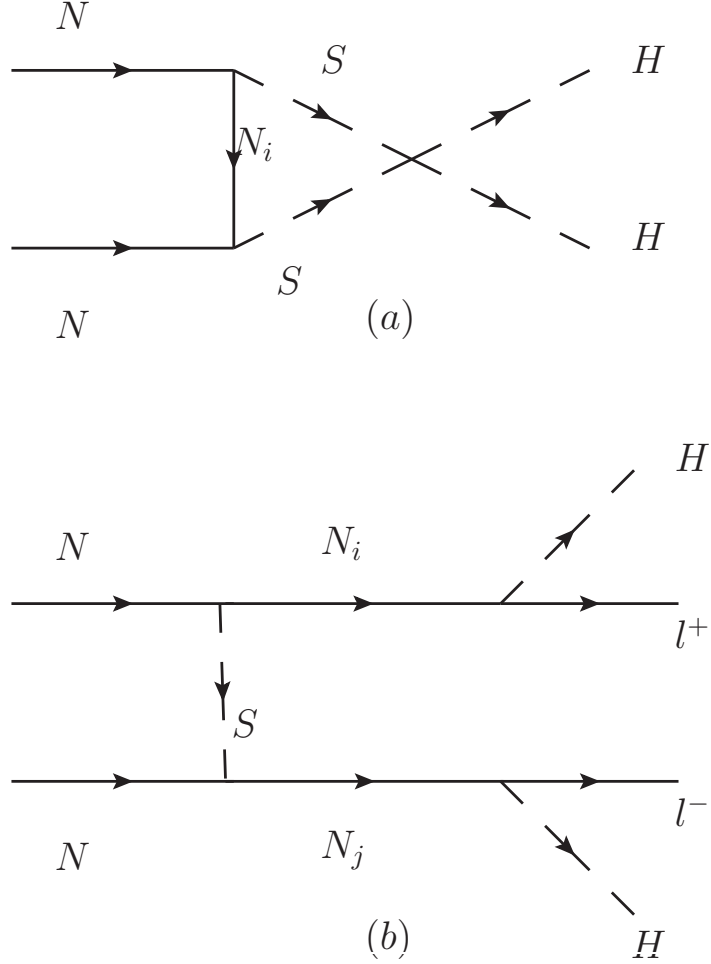}
\caption{Interactions determine the relic abundance of $N$.}
\label{fig:N as dark matter}
\end{figure}

The main interactions responsible for the relic density of $N$ are as shown in Fig. \ref{fig:N as dark matter}. Keep in mind that only $N$ and $S$ are odd under the discrete symmetry.

It is expected that $(a)$ in Fig. \ref{fig:N as dark matter}  is the dominant contribution although it is loop-suppressed since $(b)$ is of order $y^4_{\alpha i}$, which is small because we focus on the situation $m_{N_i}\leq 100$ TeV, and there is a phase suppression due to the four-body decay. Therefore, we have
\begin{equation}\label{NN to HH}
  \langle\sigma v\rangle = \frac{3}{2048 \pi^5} \frac{\lambda^4_i \lambda^2_{HS}}{m^2_{N_i}}
  \frac{T}{m_{N}}
  \left(1-\frac{m_H^2}{m_N^2}\right)^{1/2}
  \left( \int^1_0 dx \log \frac{m^2_{N_i}}{\Delta}
  \right)
  \end{equation}
where $\Delta=m^2_S+x^2 \mathrm{E}^2_{\mathrm{cm}}-x \mathrm{E}^2_{\mathrm{cm}}$ and we have assumed $m_{Ni}\gg m_S\geq m_N$. The relic abundance of $N$ is given by Eq. (\ref{relic density S}) with $S$ replaced by $N$.

 In principle, if $m_S\sim m_N$, we have to consider co-annihilation interactions, i.e., $S+N \rightarrow N_i \rightarrow l_{\alpha}^{\pm}+H^{\mp}$, whose amplitude squared is proportional to $y^2_{\alpha i}$. It is small compared to other annihilation channels because, again, we are interested in the situation of $m_{N_i}\leq 100$ TeV.

\begin{figure}
 \includegraphics[width=0.7\textwidth]{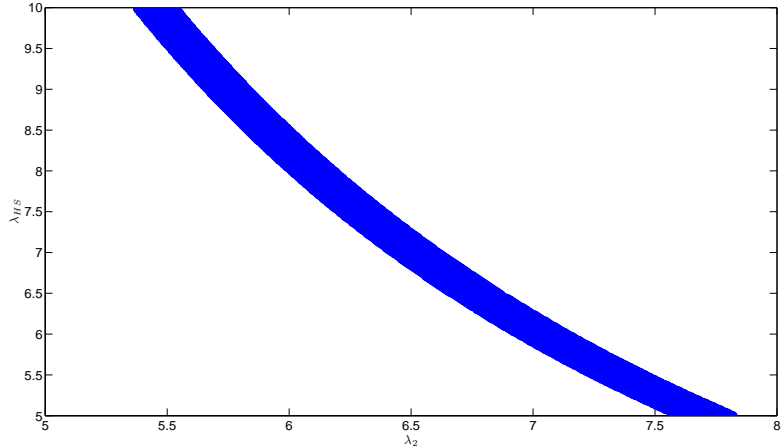}
 \caption{The blue band represents the region of the parameter space of $\lambda_2$ and $\lambda_{HS}$, which gives the right thermal relic abundance for $N_1$ and has successful low-scale leptogenesis, where $(m_H,m_N,m_S,m_{N_1},m_{N_2})=(100,150,200,3000,15000)$ in units of GeV }
 \label{fig:region for N as dark matter}
 \end{figure}

 In summary, for $S$ being the dark matter, $\langle \sigma v\rangle$ is mostly determined by $\lambda_{HS}$, which is a free parameter from the point of view of leptogenesis while for $N$, $\langle \sigma v\rangle$ is determined by $\lambda_i$ for $N_i$ running in the loop. On the other hand, with the help of the large $\lambda_2$ ($N_2$ propagating inside the loop), we can have the correct abundance for $N$. Fig. \ref{fig:region for N as dark matter} shows the allowed regions of $\lambda_2$ and $\lambda_{HS}$ for generating the right dark matter density and having low-scale leptogenesis. In this situation, we have to push both $\lambda_i$ and $\lambda_{HS}$ toward the strongly-coupled region.

\section{conclusions}\label{sec:conclusions}

In this papaer, we propose a simple and economical model, which can accommodate both leptogenesis and the dark matter, by introducing the new scalar $S$, which couples to the Higgs boson and RH Majorana neutrinos ($N_i$), and $N$. We impose a $Z_2$ symmetry under which $S$ and $N$ are odd and the rest is even; therefore, the lighter of $S$ and $N$ could be the dark matter. By increasing the coupling, $\lambda_2$, and having $N_2$ much heavier than $N_1$, we can easily increase the efficiency of generation of the lepton asymmetry without having a high reheating temperature and $N_1$ can be as low as $\mathcal{O}$(TeV). Note that all of the estimates of the lepton asymmetry are based on the one-flavor approximation instead of three flavors ($e$, $\mu$ and $\tau$). Taking into account flavor effects, the efficiency factor, $\eta'$, will be modified. From \cite{Davidson:2008bu}, $\eta'$ is enhanced by one or two orders of magnitude or remains the same order of magnitude compared to that of one-flavor approximation, which implies the estimate done before remains valid.

Finally, there have been studies of constraints on $S$ as the thermal relic via the decay of the Higgs Boson into $S$, if kinematics allows, or the elastic scattering between $S$ and nuclei, see for example, \cite{He:2008qm}.

As for $N$ as the dark matter, it can be produced in pairs via the Higgs boson and $S$. If the produced $S$ is on-shell, the decay width of $S$ into $N$ and a light neutrino is of order $\frac{m_S}{16\pi}\left(\frac{m_{\nu}}{m_{N_2}}\right)\sim 10^{-12}$ GeV, which roughly corresponds to the decay time $10^{-13}$ sec, which implies it can happen inside a detector. However, it is very challenging to identify that a new state (other than $S$) has been produced since both the $N$ and the light neutrino would escape the detector.

 At the time of writing, we notice that, in \cite{Kayser:2010fc}, they have demonstrated how one may obtain leptogenesis and neutrino mass generation in the see-saw picture without invoking a new mass scale far beyond that of electroweak symmetry breaking by introducing more than one Higgs doublet family.

\begin{acknowledgments}
I would like to thank Andr$\acute{\rm{e}}$ de Gouv$\hat{\rm{e}}$a for enlightening discussions and useful suggestions, and thank Andr$\acute{\rm{e}}$ de Gouv$\hat{\rm{e}}$a and Jennifer Kile for reading the paper. This work is supported in part by the U.S.~Department of Energy under contracts DE-AC02-06CH11357 and DE-FG02-91ER40684.
\end{acknowledgments}


\begin{thebibliography}{99}


\bibitem{McDonald:1993ex}
  J.~McDonald,
  Phys.\ Rev.\  D {\bf 50}, 3637 (1994)
  [arXiv:hep-ph/0702143].


\bibitem{neutrino masses}
B. T. Cleveland et al., Nucl. Phys. Proc. Suppl. 38, 47 (1995); Y. Fukuda et al. [Super-Kamiokande Collaboration],
Phys. Rev. Lett. 82, 2430 (1999) [arXiv:hep-ex/9812011]; K. Lande et al., Nucl. Phys. Proc. Suppl. 77, 13 (1999); D. N. Abdurashitov et al. [SAGE Collaboration], Nucl. Phys. Proc. Suppl. 77, 20 (1999); T. A. Kirsten [GALLEX and GNO Collaborations], Nucl. Phys. Proc. Suppl. 77, 26 (1999); Y. Fukuda et al. [Super-Kamiokande Collaboration], Phys. Lett. B 436, 33 (1998) [arXiv:hep-ex/9805006]; C. Athanassopoulos et al. [LSND Collaboration], Phys. Rev. C 54, 2685 (1996) [arXiv:nucl-ex/9605001]; C. Athanassopoulos et al. [LSND Collaboration], Phys. Rev. C 58, 2489 (1998) [arXiv:nucl-ex/9706006].

\bibitem{type-I seesaw}
P.~Minkowski, Phys.\ Lett.\ B {\bf 67}, 421 (1977);
M. Gell-Mann, P. Ramond and R. Slansky in {\it Supergravity},  eds. D. Freedman and P. Van Niuwenhuizen (North Holland, Amsterdam, 1979), p.~315;
T. Yanagida in {\it Proceedings of the Workshop on Unified Theory and Baryon Number in the Universe}, eds. O.~Sawada and A.~Sugamoto (KEK, Tsukuba, Japan, 1979);
S.L.~Glashow, {\it 1979 Carg\`ese Lectures in Physics -- Quarks and Leptons}, eds. M.~L\'evy {\it et al.} (Plenum, New York, 1980), p.~707;
R.N. Mohapatra and G. Senjanovi\'c, Phys.\ Rev.\ Lett.\ {\bf 44}, 912 (1980);
J.~Schechter and J.W.F.~Valle,  Phys.\ Rev.\  D {\bf 22}, 2227 (1980).



\bibitem{baryon assymetry}
    M.~Tegmark {\it et al.}  [SDSS Collaboration],
  Phys.\ Rev.\  D {\bf 69}, 103501 (2004)
  [arXiv:astro-ph/0310723];
  G.~Hinshaw {\it et al.}  [WMAP Collaboration],
  Astrophys.\ J.\ Suppl.\  {\bf 180}, 225 (2009)
  [arXiv:0803.0732 [astro-ph]].

\bibitem{first leptogenesis}
M. Fukugida and T. Yanagida, Phys. Lett. B174, (1986) 45.



\bibitem{sakharov}
A. D. Sakharov, Pisma Zh. Eksp. Teor. Fiz. 5, 32 (1967) [JETP Lett. 5, 24
(1967); Sov. Phys. Usp. 34, 392 (1991)].


\bibitem{sphaleron}
V. A. Kuzmin, V. A. Rubakov and M. A. Shaposhnikov, Phys. Lett. B155, 36 (1985);
F. R. Klinkhammer and N. S. Manton, Phys. Rev. D30, 2212 (1984).



\bibitem{Davidson:2002qv}
  S.~Davidson and A.~Ibarra,
  Phys.\ Lett.\  B {\bf 535}, 25 (2002)
  [arXiv:hep-ph/0202239].

\bibitem{Bolz:2000fu}
  M.~Bolz, A.~Brandenburg and W.~Buchmuller,
  Nucl.\ Phys.\  B {\bf 606}, 518 (2001)
  [Erratum-ibid.\  B {\bf 790}, 336 (2008)]
  [arXiv:hep-ph/0012052].

\bibitem{resonant leptogenesis}
A. Pilaftsis, Phys. Rev. D56, 5431 (1997); A. Pilaftsis and T. E. J. Underwood, Nucl. Phys. B692, 303 (2004).


\bibitem{flavor effect}
A. Abada et al. Flavour matters in leptogenesis. JHEP, 0609:010, 2006.


\bibitem{Buchmuller:2005eh}
  W.~Buchmuller, R.~D.~Peccei and T.~Yanagida,
  Ann.\ Rev.\ Nucl.\ Part.\ Sci.\  {\bf 55}, 311 (2005)
  [arXiv:hep-ph/0502169].


\bibitem{Davidson:2008bu}
  S.~Davidson, E.~Nardi and Y.~Nir,
  Phys.\ Rept.\  {\bf 466}, 105 (2008)
  [arXiv:0802.2962 [hep-ph]].

\bibitem{Petraki:2007gq}
  K.~Petraki and A.~Kusenko,
  Phys.\ Rev.\  D {\bf 77}, 065014 (2008)
  [arXiv:0711.4646 [hep-ph]].


\bibitem{DW model}
S. Dodelson and L. M. Widrow, Phys. Rev. Lett. 72, 17 (1994).


\bibitem{Law:2007jk}
C. I. Low, Phys. Rev. D 71, 073007 (2005) [arXiv:hep-ph/0501251];
  S.~S.~C.~Law and R.~R.~Volkas,
  Phys.\ Rev.\  D {\bf 75}, 043510 (2007)
  [arXiv:hep-ph/0701189].




\bibitem{McDonald:2007ka}
  J.~McDonald, N.~Sahu and U.~Sarkar,
  JCAP {\bf 0804}, 037 (2008)
  [arXiv:0711.4820 [hep-ph]].


\bibitem{Harvey:1990qw}
  S.~Y.~Khlebnikov and M.~E.~Shaposhnikov,
  Nucl.\ Phys.\  B {\bf 308} (1988) 885.
  J.~A.~Harvey and M.~S.~Turner,
  Phys.\ Rev.\  D {\bf 42}, 3344 (1990).





\bibitem{Hinshaw:2008kr}
  G.~Hinshaw {\it et al.}  [WMAP Collaboration],
  Astrophys.\ J.\ Suppl.\  {\bf 180}, 225 (2009)
  [arXiv:0803.0732 [astro-ph]].



\bibitem{f function definition}
L. Covi, E. Roulet and F. Vissani, Phys. Lett. B 384 (1996) 169.

\bibitem{Casas:2001sr}
  J.~A.~Casas and A.~Ibarra,
  Nucl.\ Phys.\  B {\bf 618}, 171 (2001)
  [arXiv:hep-ph/0103065].



\bibitem{Giudice:2003jh}
  G.~F.~Giudice, A.~Notari, M.~Raidal, A.~Riotto and A.~Strumia,
  Nucl.\ Phys.\  B {\bf 685}, 89 (2004)
  [arXiv:hep-ph/0310123].

\bibitem{Kolb:1988aj}
  E.~W.~.~Kolb and M.~S.~.~Turner,
{\it  REDWOOD CITY, USA: ADDISON-WESLEY (1988) 719 P. (FRONTIERS IN PHYSICS, 70)}




\bibitem{He:2008qm}
  X.~G.~He, T.~Li, X.~Q.~Li, J.~Tandean and H.~C.~Tsai,
  Phys.\ Rev.\  D {\bf 79}, 023521 (2009)
  [arXiv:0811.0658 [hep-ph]].

\bibitem{Kayser:2010fc}
  B.~Kayser and G.~Segre,
  arXiv:1011.6362 [hep-ph].



 \end{thebibliography}
 \end{document}